# The Bragg Diffraction Experiment Based on Ultrasonic Wave and Artificial Crystal Lattice

Qiusong Chen[*1], Wei Hou[1], Song Lin[1], GaoFu Liu[1], Weiyao Jia[†2]

1. School of Physical and Electronic Science, Guizhou Education University, Guiyang, 550018, China

2. School of Physical Science and Technology, Southwest University, Chongqing, 400715, China.

**Abstract:**

The traditional Bragg crystal diffraction experiments use X-rays, harming the participants' bodies. Therefore, many universities have not offered this basic experiment. Although microwave simulation Bragg experiments can reduce harm, there are still some potential dangers. To solve this dilemma, this article takes ultrasound as the experimental object and uses an artificial simulation of crystals to successfully achieve the Bragg crystal diffraction effect of crystals, which is in good agreement with the theoretical predictions. This experiment is expected to be widely deployed in physics, chemistry, materials, and other science and engineering majors as a basic teaching experiment.

The Bragg diffraction experiment[1] published in 1913 plays an irreplaceable role in modern materials research and is the most basic method for studying crystal structures. It has become one of the indispensable contents in physics, chemistry, materials science, and other science and engineering majors. However, since the size of crystal unit cells is generally between 0.1 and 1 nm, thus, X-rays with similar magnitude of wavelengths must be used as diffraction waves, which can cause certain harm to students and teachers in the teaching of Bragg diffraction experiments. To reduce the potentially hazardous, some teaching process of Bragg diffraction experiments utilizes microwaves[2, 3]. Since the wavelength of microwaves ranges from 1m to 1mm, artificial simulated crystals with unit cell basis vectors of approximately 3cm are mainly used for experiments. This approach not only avoids exposure to highly radioactive X-rays but also makes the experiments more intuitive and easier for students to understand.

However, the safety of microwaves remains controversial. First, the thermal effect of microwaves on biological organisms may potentially burn the experimental participants. Furthermore, there are reports that microwaves are a major potential factor for Havana syndrome[4]. Therefore, to make the experiment safer while retaining its simplicity and intuitiveness, this paper utilized ultrasonic waves as the wave source for the Bragg diffraction experiment. The 40 kHz ultrasonic waves used in this paper can effectively demonstrate the Bragg diffraction experiment in an artificial simple-cubic crystal lattice with a unit cell basis vector of 14 mm. The experimental results are in good agreement with the theoretical analysis of (100) and (110) crystal planes. This method can not only be used as a basic experiment in university physics, avoiding the harmful radiation effects of traditional X-rays and microwaves but also provides an effective means for analyzing the physical properties of ultrasonic waves.

---

[*]Corresponding author. E-mail addresses: chenqiusong@gmail.com, qschen16@fudan.edu.cn
[†]Corresponding author. E-mail addresses: wyjia@swu.edu.cn

## 1、The Principle of Bragg Diffraction:

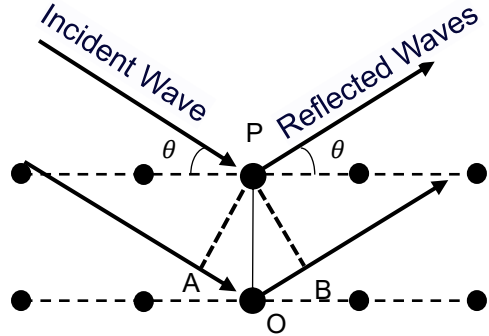

Figure 1: The schematic diagram of crystal Bragg diffraction

As shown in Figure 1, when two parallel incident waves reflect from two lattice points O and P on the same set of crystal planes, if the path difference between the two reflected waves satisfies the Bragg formula, diffraction will be enhanced:

$$2d \sin \theta = n\lambda \tag{1}$$

where $d$ is the interplanar spacing, $\theta$ is the angle between the incident and reflected waves and the crystal plane, $n$ is the interference order, and $\lambda$ is the wavelength.

For a simple cubic crystal, the interplanar spacing is determined by the Miller indices:

$$d = \frac{a}{\sqrt{h^2+k^2+l^2}} \tag{2}$$

where $a$ is the basis vector length of the crystal, and $h$、$k$、$l$ are the Miller indices for a specific set of crystal planes.

Therefore, by measuring the relationship between diffraction intensity and angle, we can analyze the crystal structure, or conversely, measure the characteristics of the diffracted waves while given the known crystal structure.

## 2、Apparatus:

The standard Bragg experiment, which is designed to match the interplanar spacing of crystals (approximately 0.1 nm), can only observe the diffraction effect from X-rays in crystals. However, in this experiment, ultrasonic waves will be used as the source for investigation. To reduce experimental costs, piezoelectric transducers from commercial ultrasonic ranging modules were employed as the ultrasonic wave generation and reception devices, as shown in Figure 2(a). To ensure that the diffraction of ultrasonic waves in the crystal more closely matches the requirements of Equation (1), the incident waves should be plane waves. Therefore, the ultrasonic wave generator will be arranged in a 3×3 matrix to generate the plane waves.

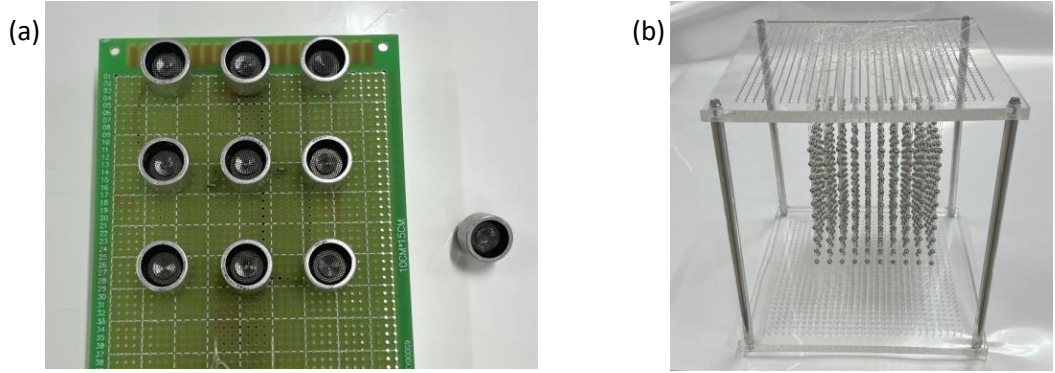

Figure 2: (a) The ultrasonic transmitting(left) and receiving devices(right), (b) The artificial crystalline.

A function generator (Victory VC2015H) was employed to drive the ultrasonic generator, exciting a sinusoidal wave at 40 kHz with an amplitude of 0.5 V. A lock-in amplifier (Sine Scientific Instruments OE1022) is used to measure the signals collected by the ultrasonic receiver.

The artificial simple cubic crystal lattice used in this experiment is shown in Figure 2(a). The metal spheres arranged as a matrix of 9×10×10 were used to imitate the atomic kernel. They are made of stainless steel, which has a high acoustic reflection coefficient[5], while the connecting lines are made of silicone tubes with a lower reflection coefficient. Since the maximum response frequency of the piezoelectric transducer is 40 kHz, and the speed of sound in air is 340 m/s, the corresponding ultrasonic wavelength is 8.6 mm. Therefore, 14 mm is chosen as the basis vector length of the artificial simulated crystal.

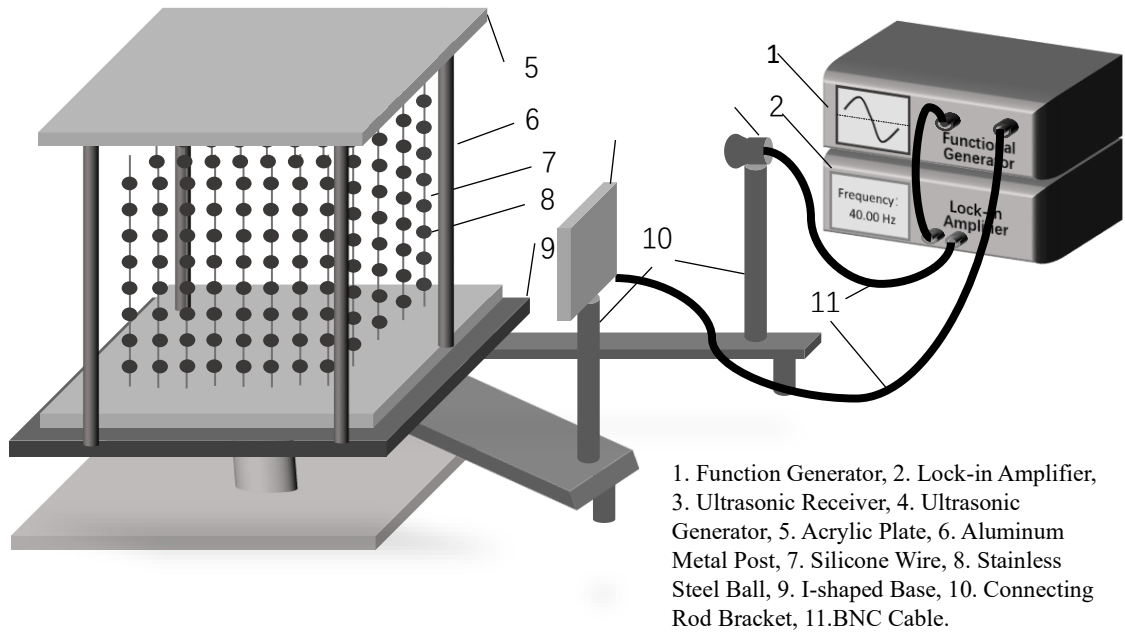

1. Function Generator, 2. Lock-in Amplifier, 3. Ultrasonic Receiver, 4. Ultrasonic Generator, 5. Acrylic Plate, 6. Aluminum Metal Post, 7. Silicone Wire, 8. Stainless Steel Ball, 9. I-shaped Base, 10. Connecting Rod Bracket, 11.BNC Cable.

Figure 3. The schematic diagram of experimental apparatus

## 3、Experimental Process and Result Analysis:

As shown in Figure 3, connect all the instruments and place the "I-shaped" base at the center of the semicircular desktop with angle scales. Meanwhile, align the base vector direction of the simulated crystal with the diameter of the semicircle, and ensure that the center of the crystal also coincides with the center of the semicircle in the vertical direction, as illustrated in Figure 4(a).

During the measurement, adjust the connecting rod between the emitter and receiver to maintain the same angle with the diameter. Then, vary the included angle $\theta$ between 10° and 70°, and use the lock-in amplifier to measure the signal intensity at the receiver for every 2° interval.

### 3.1、Bragg Diffraction of (100) Crystal Plane

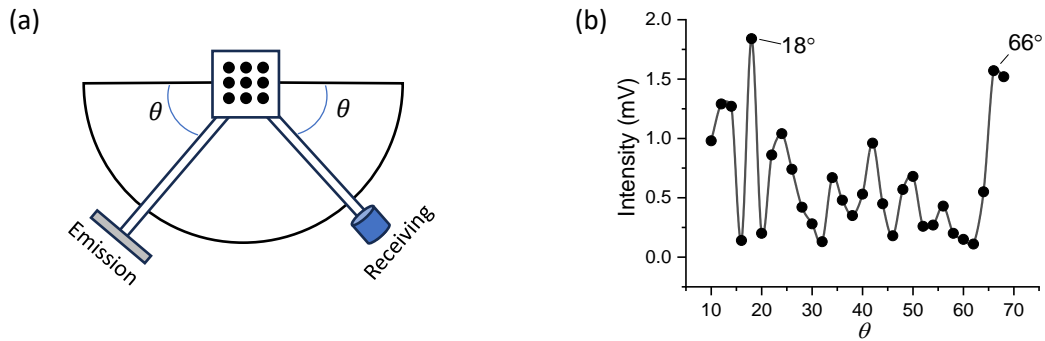

Figure 4 (a) The measurement method of (100) crystal plane. (b) The curve of diffraction intensity versus diffraction angle.

Figure 4(b) is the diffraction result of the (100) crystal plane, it can be observed that the reflected wave is relatively strong at 10°, which is due to the incident wave directly reaching the receiver. As the angle increases, the diffraction wave intensity begins to decrease, but there is a sharp peak at 18° followed by fluctuations and another sharp peak at 66°. According to Equations (1) and (2), there are three diffraction enhancement peaks for the (100) crystal plane, which are at 17.89°, 37.90°, and 67.14° respectively. However, in Figure 4(b), only two diffraction enhancement peaks are observed at 18° and 66°, corresponding to the first and third diffraction peaks of the (100) crystal plane. The second diffraction peak at 37.90° is not present. This is due to the four aluminum pillars located at the corners of the artificial crystal simulator (see Figure 2(b)). These pillars block the incident and diffracted waves in the direction corresponding to the diffraction angle of around 37.90°. To avoid this limitation, the next step will be to measure the diffraction of the (110) crystal plane using the gaps between the four aluminum pillars.

### 3.2、Bragg Diffraction of (110) Crystal Plane

The direction of the simulated crystal was altered to align the (110) crystal direction parallel to the diameter of the semicircle on the desktop, with the center remaining directly above the center of the circle, as illustrated in Figure 5(a). Then, while varying the included angle $\theta$ between 10° and 70°, the signal intensity at the receiver end was measured using the lock-in amplifier every 2° interval. The experimental results are shown in Figure 5(b).

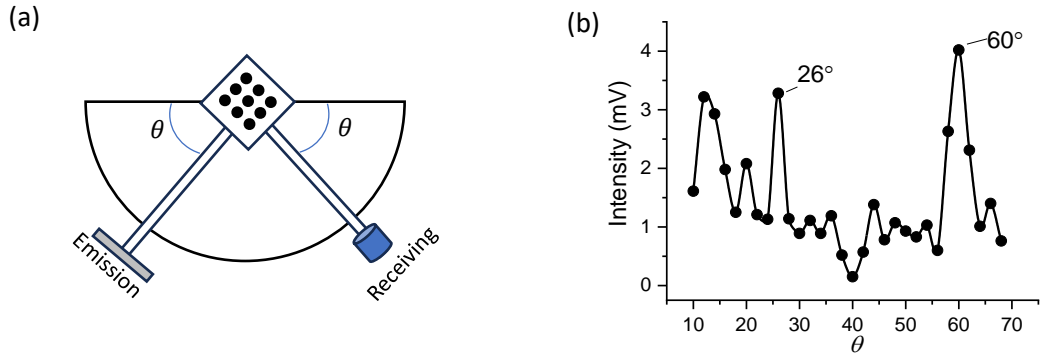

Figure 5 (a) The measurement method of (110) crystal plane and (b) the curve of diffraction intensity versus diffraction angle.

As seen in Figure 5(b), the reflected wave is relatively strong at 12°. As the angle increases, the diffraction wave intensity begins to decrease, but there is a sharp peak at 26°, followed by fluctuations and another sharp peak at 60°. According to Equations (1) and (2), there are two diffraction enhancement peaks for the (110) crystal plane, which are at 25.74° and 60.31°, precisely matching the experimental results.

## Conclusion

In this study, the use of planar ultrasonic waves generated by piezoelectric transducers successfully achieved the Bragg diffraction effect in artificial crystals, presenting good agreement between theoretical analysis and experimental results. This experimental method avoids potential harmful factors associated with traditional Bragg diffraction experiments using X-rays and microwaves, while also providing a visual demonstration of the diffraction effect of crystal structure on waves. Therefore, it can be used as a basic experiment for undergraduate physics and materials science courses in universities. Additionally, this experiment also provides a convenient experimental method for measuring and analyzing the physical parameters of ultrasonic waves.

## Acknowledgments

This work was financially supported by Guizhou Provincial Science and Technology Projects (Grant No: QKHJC-ZK[2021]YB329 & QKHJC-ZK [2021]YB018), the undergraduate teaching quality and teaching reform project "Reform and Practice Research of College Physics Experiment Teaching Mode" in Guizhou Province in 2022, and Chongqing graduate education teaching reform research project (Grant No: yjg223035)